\documentclass[aps,prd,twocolumn,superscriptaddress,nofootinbib]{revtex4-2}
\usepackage{amsmath, amssymb, bm}  
\usepackage{graphicx}  
\usepackage{tensor} 
\usepackage{siunitx}
\usepackage[colorlinks=true,linkcolor=blue,citecolor=blue,urlcolor=blue]{hyperref} 
\usepackage{lineno}
\usepackage{tikz}
\usepackage{orcidlink}
\begin{document}
\title{Search for Long-Transient Gravitational Waves from Supernova SN2023ixf\\ using GFH-v2 Pipeline}
\author{Sandhya Sajith Menon\,\orcidlink{0009-0008-4985-1320}}
\affiliation{Physics Department, Ariel University, Ariel, Israel}
\affiliation{Istituto Nazionale di Fisica Nucleare-Roma 1, Piazzale Aldo Moro 2, I-00185, Rome, Italy}
\affiliation{Università di Roma La Sapienza, I-00185 Rome, Italy}

\author{Lorenzo Pierini\,\orcidlink{0000-0003-0945-2196}}
\affiliation{Istituto Nazionale di Fisica Nucleare-Roma 1, Piazzale Aldo Moro 2, I-00185, Rome, Italy}

\author{Pia Astone\,\orcidlink{0000-0003-4981-4120}}
\affiliation{Istituto Nazionale di Fisica Nucleare-Roma 1, Piazzale Aldo Moro 2, I-00185, Rome, Italy}

\author{Cristiano Palomba\,\orcidlink{0000-0002-4450-9883}}
\affiliation{Istituto Nazionale di Fisica Nucleare-Roma 1, Piazzale Aldo Moro 2, I-00185, Rome, Italy}

\author{Francesco \surname{Safai Tehrani}\,\orcidlink{0000-0001-7796-0120}}
\affiliation{Istituto Nazionale di Fisica Nucleare-Roma 1, Piazzale Aldo Moro 2, I-00185, Rome, Italy}

\author{Lorenzo Silvestri\,\orcidlink{0009-0008-5207-661X}}
\affiliation{Università di Roma La Sapienza, I-00185 Rome, Italy}
\affiliation{INFN-CNAF - Bologna, Viale Carlo Berti Pichat, 6/2, 40127 Bologna BO, Italy}

\author{Ornella Juliana Piccinni\,\orcidlink{0000-0001-5478-3950}}
\affiliation{IAC3, Universitat de les Illes Balears, Cra. de Valldemossa km 7.5, 07122 Palma, Spain}

\author{Simone Dall’Osso\,\orcidlink{0000-0003-4366-8265}}
\affiliation{Dipartimento di Fisica e Astronomia, Alma Mater Studiorum - Università di Bologna, I-40127 Bologna, Italy}

\author{Stefano \surname{Dal Pra}\,\orcidlink{0000-0002-1057-2307}}
\affiliation{INFN-CNAF - Bologna, Viale Carlo Berti Pichat, 6/2, 40127 Bologna BO, Italy}

\author{Sabrina \surname{D'Antonio}\,\orcidlink{0000-0003-0898-6030}}
\affiliation{Istituto Nazionale di Fisica Nucleare-Roma 1, Piazzale Aldo Moro 2, I-00185, Rome, Italy}

\author{Sergio \surname{Frasca}\,\orcidlink{0000-0002-0898-6703}}
\affiliation{Istituto Nazionale di Fisica Nucleare-Roma 1, Piazzale Aldo Moro 2, I-00185, Rome, Italy}

\author{Dafne Guetta\,\orcidlink{0000-0002-7349-1109}}
\affiliation{Physics Department, Ariel University, Ariel, Israel}

\author{Paola Leaci\,\orcidlink{0000-0002-3997-5046}}
\affiliation{Università di Roma La Sapienza, I-00185 Rome, Italy}
\affiliation{Istituto Nazionale di Fisica Nucleare-Roma 1, Piazzale Aldo Moro 2, I-00185, Rome, Italy}

\author{Michał Bejger\,\orcidlink{0000-0002-4991-8213}}
\affiliation{Nicolaus Copernicus Astronomical Center, Polish Academy of Sciences, 00-716 Warsaw, Poland}
\affiliation{INFN Sezione di Ferrara, Via Saragat 1, 44122 Ferrara, Italy}

\author{Alicia Calafat\,\orcidlink{0009-0008-7515-6305}}
\affiliation{IAC3, Universitat de les Illes Balears, Cra. de Valldemossa km 7.5, 07122 Palma, Spain}

\author{Evan Goetz\,\orcidlink{0000-0003-2666-721X}}
\affiliation{University of British Columbia}

\author{Rafel Jaume Amengual\,\orcidlink{0000-0001-8691-3166}}
\affiliation{IAC3, Universitat de les Illes Balears, Cra. de Valldemossa km 7.5, 07122 Palma, Spain}

\author{Ian Jones}
\affiliation{University of Southampton, Southampton SO17 1BJ, United Kingdom}

\author{David Keitel\,\orcidlink{0000-0002-2824-626X}}
\affiliation{IAC3, Universitat de les Illes Balears, Cra. de Valldemossa km 7.5, 07122 Palma, Spain}

\author{Andrzej Królak\,\orcidlink{0000-0003-4514-7690}}
\affiliation{Institute of Mathematics, Polish Academy of Sciences, Warsaw, Poland}
\affiliation{National Centre for Nuclear Research, Swierk, Poland}

\author{Iuri La Rosa\,\orcidlink{0000-0003-0107-1540}}
\affiliation{IAC3, Universitat de les Illes Balears, Cra. de Valldemossa km 7.5, 07122 Palma, Spain}

\author{Andrew Melatos\,\orcidlink{0000-0003-4642-141X}}
\affiliation{School of Physics, University of Melbourne, Parkville VIC 3010, Australia}

\author{Joan-René Mérou\,\orcidlink{0000-0002-5776-6643}}
\affiliation{IAC3, Universitat de les Illes Balears, Cra. de Valldemossa km 7.5, 07122 Palma, Spain}

\author{Lorenzo Mirasola\,\orcidlink{0009-0004-0174-1377}}
\affiliation{IAC3, Universitat de les Illes Balears, Cra. de Valldemossa km 7.5, 07122 Palma, Spain}

\author{Claudio Salvadore\,\orcidlink{0009-0002-9967-4111}}
\affiliation{Università di Roma La Sapienza, I-00185 Rome, Italy}
\affiliation{Istituto Nazionale di Fisica Nucleare-Roma 1, Piazzale Aldo Moro 2, I-00185, Rome, Italy}

\author{Alicia M. Sintes\,\orcidlink{0000-0001-9050-7515}}
\affiliation{IAC3, Universitat de les Illes Balears, Cra. de Valldemossa km 7.5, 07122 Palma, Spain}

\author{Karl Wette\,\orcidlink{0000-0002-4394-7179}}
\affiliation{OzGrav-ANU, Centre for Gravitational Astrophysics, Australian National University, Canberra ACT 2601, Australia}

\date{\today}

\begin{abstract}
We present a directed search for long-transient gravitational waves from the possible newborn magnetar remnant of SN~2023ixf, a nearby Type II core-collapse supernova in the M101 galaxy. The analysis uses LIGO Hanford and Livingston data from Engineering Run 15, using coincident data lying within the on-source window associated with the supernova. We target signals from a rapidly rotating, non-axisymmetric neutron star whose spin-down is dominated by gravitational-wave emission, producing a power-law decrease in frequency and a corresponding decrease in strain amplitude. The search is performed with the GFH-v2 pipeline, based on the Generalized Frequency Hough transform. No candidate survives the coincidence and follow-up analysis. We therefore set upper limits on the maximum detectable distance as a function of initial frequency and ellipticity. For the highest ellipticity interval, the 90\% upper limits reach distances of about $1$-$2.5$~Mpc across most of the analysed band. Although these limits are below the distance to M101, the search provides the first application of GFH-v2 to a nearby core-collapse supernova and characterizes its performance on real detector data.
\end{abstract}

\maketitle

\section{Introduction}

Rapidly rotating newborn magnetars formed in core-collapse supernovae (CCSNe) or binary neutron star (BNS) mergers are promising sources of long-transient gravitational waves (tCWs). These signals are different from the short-duration gravitational-wave bursts that may be produced directly during the core-collapse process~\cite{2012ARNPS62407J,2016PhRvD93d2002G,Abdikamalov2022,2009CQGra26f3001O}. Burst signals are associated with the rapid dynamics of the collapse itself whereas the signals considered in this work are emitted after the collapse by a newly formed, rapidly rotating neutron star or magnetar and may last from several hours to days\cite{2015PASA3234L,2023LRR263R,2011PhRvD83j4014C,2001A&A367525P,2002PhRvD66h4025C,2009ApJ7021171C,2021MNRAS5024680S,2022ASSL465245D}. If the newly formed magnetar possesses a non-axisymmetric mass distribution, gravitational waves are emitted as rotational energy is lost through spin-down \cite{2009MNRAS3981869D,2012ApJ76163P,dallo15,2018MNRAS4801353D,2020MNRAS4944838L}. Such non-axisymmetric deformations are expected to arise primarily from the extremely strong internal magnetic fields of newborn magnetars, which can induce large ellipticities and sustain the time-varying mass quadrupole responsible for tCW emission. Other mechanisms, such as crustal deformations or unstable oscillation modes, may also contribute, although they are generally expected to produce smaller quadrupole deformations~\cite{1995ApJ442259L,2000MNRAS319902U,2008MNRAS385531H}. Detection of these signals would provide direct information on the birth and early evolution of neutron stars, their internal structure, and the physical mechanisms governing their spin evolution.

tCWs occupy an intermediate regime between short-duration compact binary coalescence signals and nearly monochromatic continuous waves. Their frequency decreases continuously as the source spins down, requiring search methods capable of following a rapidly evolving signal. Because the signal parameters are generally unknown and the parameter space is large, searches for long-transient signals remain computationally challenging. Hierarchical semi-coherent approaches provide an efficient solution by balancing computational cost and sensitivity while exploring broad regions of parameter space~\cite{2018PhRvD98j2004M,2019PhRvD99j4067O,2019PhRvD99l3003S,2019PhRvD100f2005M,2019PhRvD100b4034B,2023PhRvD108l3045G,2026PhRvD113l3042S}. In particular, the method used in this work is the Generalised Frequency Hough  transform~\cite{2018PhRvD98j2004M, 2026PhRvD113l3042S}.

SN2023ixf is a nearby Type II core-collapse supernova and is therefore an interesting target for a directed search for this type of long-transient signal. Discovered on 19 May 2023 in the nearby galaxy M101 at a distance of approximately 6.8~Mpc, it is one of the closest core-collapse supernovae observed in recent decades~\cite{2023TNSTR11581I,2022ApJ934L7R}. If the explosion produced a rapidly rotating magnetar, the source could emit tCWs during the first hours or days following core collapse. The epoch of SN~2023ixf overlaps LIGO Engineering Run 15 (ER15)~\cite{Abac_2026,GWOSC,2015CQGra32g4001L}, providing an opportunity to perform a directed search using real interferometric data obtained immediately before the start of the fourth observing run (O4).

In previous work~\cite{2026PhRvD113l3042S}, we developed and validated the improved version of Generalized Frequency Hough pipeline (GFH-v2) for tCW searches using extensive signal injection studies in O4a data. In the present work, we report its first application to a directed search for tCW emission from the possible compact remnant of SN~2023ixf. We analyse coincident data from the LIGO Hanford (H1) and Livingston (L1) detectors~\cite{2015CQGra32g4001L,2025PhRvD111f2002C}, perform candidate selection and follow-up, and determine upper limits on the maximum detectable distance as a function of the source parameters. Although no gravitational-wave signal is identified, this work demonstrates the performance of the GFH-v2 pipeline under realistic detector conditions and establishes a framework for future searches targeting nearby magnetar-forming events.

The paper is structured as follows: Section~\ref{sec:event_model} introduces SN~2023ixf and the gravitational-wave signal model adopted in this work. Section~\ref{sec:data_method} describes the ER15 dataset and the GFH-v2 search pipeline. The search configuration, explored parameter space, critical ratio estimation, and computational implementation are presented in Section~\ref{sec:search_config}. The search results and the candidate follow-up procedure are reported in Sections~\ref{sec:results} and \ref{sec:followup}, respectively. The upper-limit estimation is presented in Section~\ref{sec:upperlimits}. Finally, the conclusions are summarized in Section~\ref{sec:conclusions}.

\section{SN2023ixf and Signal Model}
\label{sec:event_model}
\subsection{SN~2023ixf}

SN~2023ixf was discovered on 19 May 2023 by the amateur astronomer Kōichi Itagaki in the nearby spiral galaxy M101 (the Pinwheel Galaxy)~\cite{2023TNSTR11581I}. It was subsequently classified as a Type II core-collapse supernova based on its spectral features~\cite{2023TNSAN1191P}. Early-time spectroscopy revealed strong hydrogen emission together with narrow flash-ionization features, indicating interaction between the expanding ejecta and dense circumstellar material (CSM) surrounding the progenitor~\cite{2023ApJ956L5B}.

The supernova is located at right ascension $\alpha = 14^\mathrm{h}03^\mathrm{m}38.580^\mathrm{s}$ and declination $\delta = +54^\circ18'42.10''$. For the present analysis, we adopt a distance of $6.85$~Mpc to M101, based on the Cepheid distance measurement of Ref.~\cite{2022ApJ934L7R}. SN~2023ixf is therefore one of the nearest core-collapse supernovae observed in recent decades. Because of its proximity, the event became an important target for multi-messenger observations, including electromagnetic, neutrino, and gravitational-wave searches~\cite{2025ApJ985183A,2023ApJ955L9G}.

Observations across the electromagnetic spectrum indicate that the progenitor was a red supergiant with a dense but spatially confined circumstellar environment produced by enhanced mass loss before explosion~\cite{2025Univ11231J,2023ApJ95646S}. Such progenitors may leave behind a rapidly rotating neutron star or magnetar capable of emitting tCWs. Moreover, the early electromagnetic observations constrain the explosion epoch to within about two hours, providing an accurate trigger time for a directed gravitational-wave search.

Table~\ref{tab:SN_properties} summarizes the main properties of SN~2023ixf used in this work.

\begin{table}[h]
\centering
\begin{tabular}{lc}
\hline
Property & Value \\
\hline
Discovery date & 19 May 2023~\cite{2023TNSTR11581I} \\
Type & Type II core-collapse supernova \\
Host galaxy & M101 (Pinwheel Galaxy) \\
Distance & $\sim$6.85 Mpc \\
Coordinates & RA=210.91075°, DEC=54.311694° \\
Progenitor mass & $\sim$8-15 M$_\odot$ \\
EM counterpart & Optical, X-ray, radio emission \\
\hline
\end{tabular}
\caption{Main properties of SN~2023ixf used in this work~\cite{2023TNSTR11581I,2023TNSAN1191P,2025Univ11231J,2025ApJ985183A,2023ApJ955L9G}.}
\label{tab:SN_properties}
\end{table}

\subsection{Gravitational-wave signal model}

In this work we search for tCWs from a rapidly rotating newborn magnetar. We assume that the neutron star rotates about one of its principal axes and possesses a non-axisymmetric deformation characterized by an ellipticity $\epsilon$. Such a deformation gives rise to gravitational-wave emission as the star spins down.

The gravitational-wave strain amplitude is given by~\cite{1998PhRvD58f3001J}
\begin{equation}
h_0(t)=\frac{4\pi^2G}{c^4}\frac{I\epsilon}{d}f_{\rm gw}^2(t),
\label{eq:amplitude}
\end{equation}
where $I$ is the moment of inertia of the neutron star, $d$ is the source distance, and $f_{\rm gw}(t)$ is the instantaneous gravitational-wave frequency.
\begin{table*}[htbp]
\centering
\begin{tabular}{cllccr}
\hline
Segment & UTC start & UTC end & GPS start & GPS end & Duration \\
\hline
1 & 2023-05-17 09:25:20 & 2023-05-17 15:06:40 
  & 1368350720 & 1368371200 & 20480\,s ($\approx$5.7\,hr) \\
2 & 2023-05-18 04:19:26 & 2023-05-18 09:10:39 
  & 1368418766 & 1368436239 & 17473\,s ($\approx$4.9\,hr) \\
3 & 2023-05-18 12:57:15 & 2023-05-18 15:55:01 
  & 1368449835 & 1368460501 & 10666\,s ($\approx$3.0\,hr) \\
\hline
\multicolumn{5}{l}{Total analysed duration} 
  & 48\,619\,s ($\approx$0.56\,days) \\
\hline
\end{tabular}
\caption{Coincident H1-L1 science segments from ER15 used in this search. The selected segments provide coverage around the estimated explosion epoch of SN~2023ixf. GPS times are given for reference, together with the corresponding UTC intervals and analyzed durations.}
\label{tab:segments}
\end{table*}
The rotational evolution of the star is described by a power-law spin-down,
\begin{equation}
\dot{f}_{\rm rot}\propto-f_{\rm rot}^{n},
\label{eq:spindown}
\end{equation}
where $n$ is the braking index, whose value depends on the dominant energy-loss mechanism~\cite{2015PhRvD91f3007H}. For a rigid, non-precessing neutron star rotating about a principal axis, the gravitational-wave frequency is twice the rotational frequency~\cite{Maggiore:2007ulw},
\begin{equation}
f_{\rm gw}=2f_{\rm rot}.
\label{eq:fgw_frot}
\end{equation}

The gravitational-wave frequency evolution can therefore be written as
\begin{equation}
\dot{f}_{\rm gw}=-k f_{\rm gw}^{n},
\label{eq:spindown_gw}
\end{equation}
which has the solution
\begin{equation}
f_{\rm gw}(t)=
\frac{f_{\rm gw,0}}
{\left(1+\frac{t-t_0}{\tau}\right)^{1/(n-1)}},
\label{eq:frequency}
\end{equation}
where $f_{\rm gw,0}$ is the gravitational-wave frequency at the reference time $t_0$, and
\begin{equation}
\tau=\frac{1}{(n-1)k f_{\rm gw,0}^{n-1}}
\end{equation}
is the characteristic spin-down timescale. For simplicity, in the following we denote the gravitational-wave
frequency $f_{\rm gw}$ by $f$, its time derivative
$\dot{f}_{\rm gw}$ by $\dot{f}$, and its initial value
$f_{\rm gw,0}$ by $f_0$.

In this search we assume that the spin-down is dominated by gravitational-wave emission, corresponding to a braking index $n=5$~\cite{Shapiro:1983du}. In this case,
\begin{equation}
k=\frac{32\pi^4G}{5c^5}\epsilon^2I.
\end{equation}

Substituting Eq.~(\ref{eq:frequency}) into Eq.~(\ref{eq:amplitude}) gives the time evolution of the signal amplitude,
\begin{equation}
h_0(t)=
\frac{4\pi^2G}{c^4}
\frac{I\epsilon}{d}
f_0^2
\left(1+\frac{t-t_0}{\tau}\right)^{-1/2}.
\label{eq:amplitude_evolved}
\end{equation}

This signal model defines the waveform targeted throughout the analysis.

\section{Data and Search Method}
\label{sec:data_method}
\subsection{ER15 data}
\label{ER15data}

The search is performed using data from Engineering Run 15 (ER15)~\cite{Abac_2026}, a commissioning run of the Advanced LIGO detectors at the Hanford and Livingston observatories~\cite{2015CQGra32g4001L,2025PhRvD111f2002C} carried out between 26 April and 24 May 2023, immediately before the start of the fourth observing run (O4)~\cite{2025CQGra42h5016S}. During ER15, the detectors were operated under near-observing conditions while commissioning, calibration, and detector characterization activities were still ongoing. Although engineering runs are primarily intended to prepare the instruments for science observations, portions of the data satisfy the quality requirements needed for astrophysical analyses.

The on-source window associated with SN~2023ixf overlaps ER15, making these data suitable for a search associated with the event. Virgo~\cite{2015CQGra32b4001A} was not operating during this period and therefore does not contribute to the present analysis. We use the subset of ER15 data approved by the LIGO-Virgo-KAGRA Collaboration~\cite{2020LRR233A,2015CQGra32g4001L,2015CQGra32b4001A,2021PTEP2021eA102A} for studies of SN~2023ixf. The first light from SN~2023ixf is estimated to have occurred at MJD $60082.743 \pm 0.083$ (2023 May 18 $17{:}50 \pm 2$\,hr UTC)~\cite{2023ApJ955L8H}. The core collapse and formation of a compact remnant are expected to precede the emergence of the electromagnetic emission by a model-dependent delay associated with shock propagation through the progenitor and its circumstellar environment. We therefore analyse the available coincident H1-L1 science segments within the conservative on-source window adopted for SN~2023ixf~\cite{T2400184,2025ApJ985183A}. Under the hypothesis that core collapse occurred during the available coincident data, these segments provide an opportunity to search for tCW emission from a possible newborn neutron star.

\begin{figure}[htbp]
\centering
\includegraphics[width=0.45\textwidth]{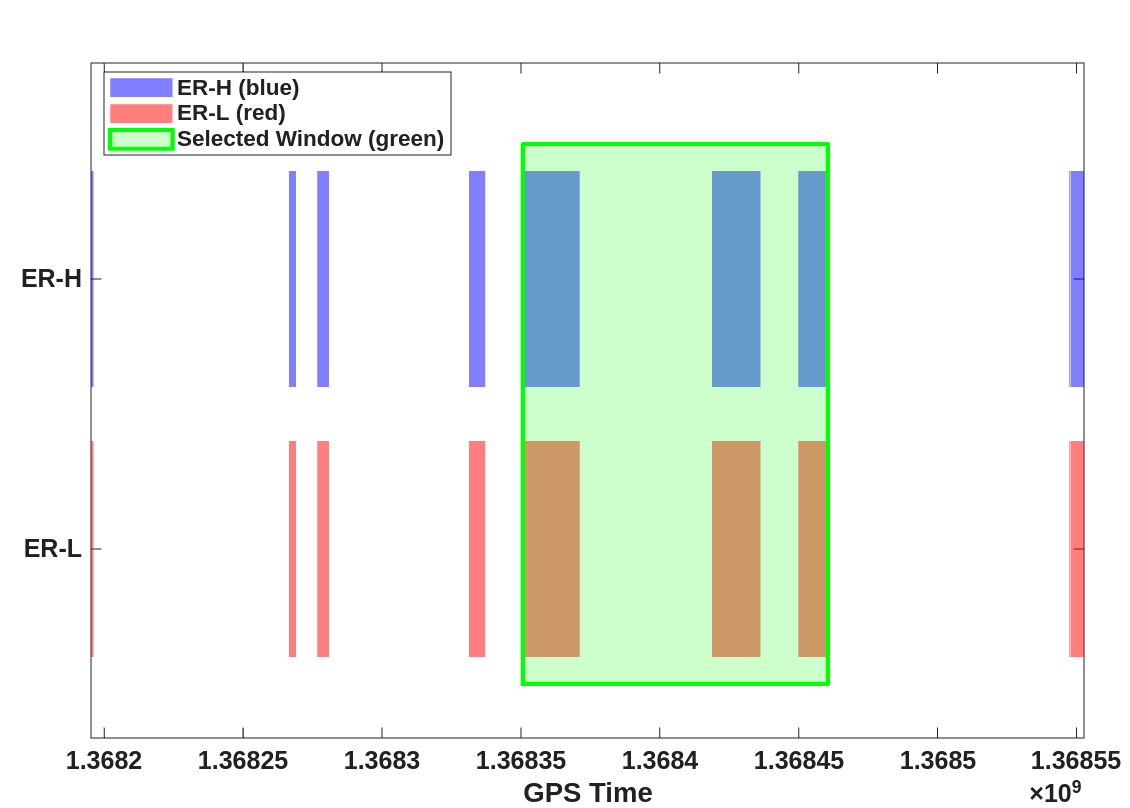}
\caption{Coincident H1-L1 science segments from ER15 used in this search. The selected segments lie within the conservative on-source window adopted for SN~2023ixf~\cite{2025ApJ985183A}.}
\label{fig:ERscience}
\end{figure}

The final dataset consists of three coincident H1-L1 science segments, summarised in Table~\ref{tab:segments}. Segments occurring well before the explosion window and two short segments ($<500$\,s each) following a long gap after the main dataset are excluded as they do not contribute meaningfully to a long-transient analysis. The total analysable coincident data amounts to 48\,619\,s ($\approx 0.56$\,days); including gaps between segments, the analysis window spans approximately 
1.27 days, allowing signals with durations up to this window to be tracked. Their distribution within the analysis window is shown in Fig.~\ref{fig:ERscience}.

\begin{figure}[htbp]
    \centering
    \includegraphics[width=0.9\linewidth]{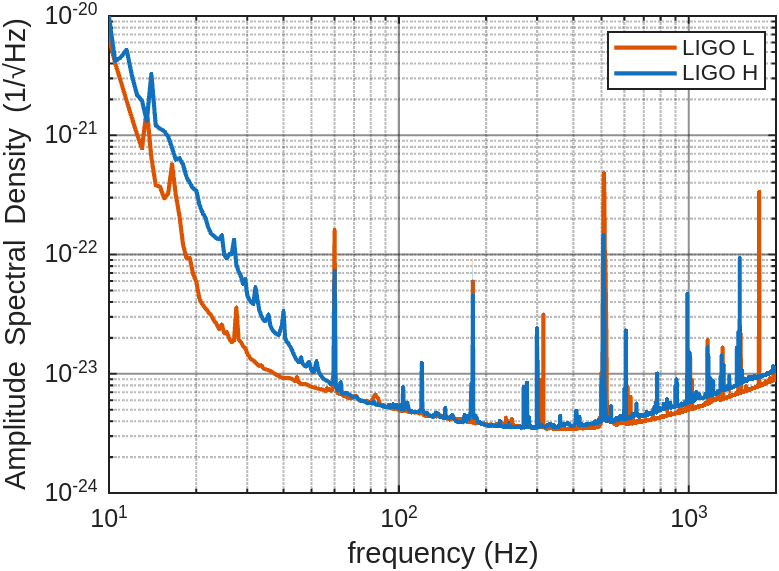}
    \caption{
    Amplitude spectral densities of the LIGO H1 and L1 detectors computed from the approved ER15 SFDB dataset.}
    \label{fig:ASDER}
\end{figure}

Figure~\ref{fig:ASDER} compares the H1 and L1 amplitude spectral densities (ASDs) obtained from the approved ER15 SFDB dataset. These ASDs are computed from the full approved ER15 SFDB dataset~\cite{Astone2025SFDB}. For each detector, the ASD is obtained by averaging the power spectra of the individual SFDB FFT segments belonging to science time, after applying the basic data-quality vetoes. A weighted averaging procedure is used to reduce the influence of noisier spectra, following the same convention adopted in Frequency-Hough-based searches~\cite{2014PhRvD90d2002A}. The square root of the resulting averaged power spectral density is then
taken to obtain the ASD. The two detectors exhibit comparable sensitivity over the analysed frequency range, providing a consistent two-detector dataset for the directed search.

\begin{figure}[htbp]
    \centering
    \includegraphics[width=0.9\linewidth]{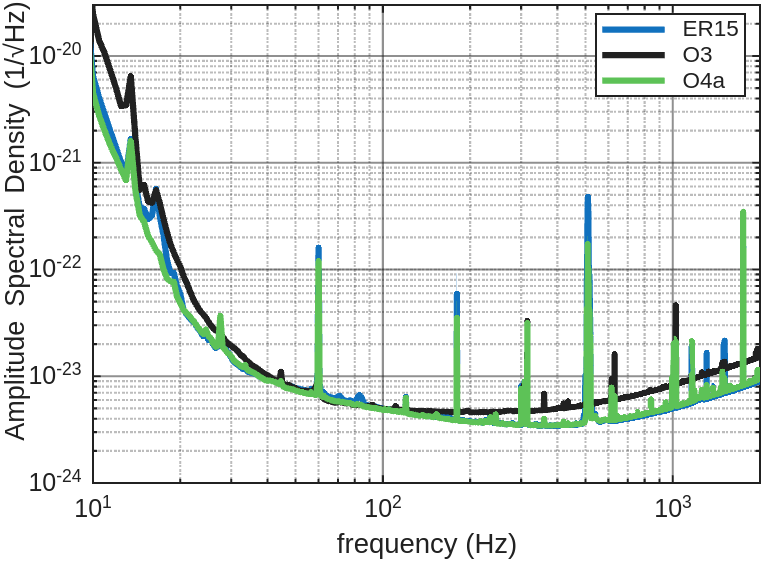}
    \caption{
    Comparison of the Livingston amplitude spectral density during ER15 with representative spectra from the O3 and O4a observing runs. }
    \label{fig:ASDcomparison}
\end{figure}

To assess the suitability of ER15 for this search, we compare this with representative ASDs from the O3~\cite{2023ApJS26729A} and O4~\cite{2025arXiv250818079T} observing runs. As shown in Fig.~\ref{fig:ASDcomparison}, the ER15 sensitivity is noticeably better than O3 and is close to that achieved during O4a over most of the analysed frequency range of 600-2000\,Hz. The overall spectral shape is very similar to O4a, although some excess noise is present at a few frequencies. Despite these local features, the comparison shows that the ER15 data have sufficient sensitivity to perform a meaningful search for tCWs associated with SN~2023ixf.

\subsection{Search method: GFH-v2}
\label{subsec:gfhv2}

The search is performed with the GFH-v2 pipeline, which is designed for tCW signals with a power-law frequency evolution. A complete description of the method, including the implementation and validation on O4a data, is given in~\cite{2026PhRvD113l3042S}. Here we summarize the main elements needed for the SN~2023ixf search.

The input data are taken from the Short Fourier Transform Database (SFDB)~\cite{2005CQGra22S1197A,Astone2025SFDB}. For each combination of initial frequency $f_0$ and ellipticity $\epsilon$, the frequency band expected to contain the signal is selected from the SFDB and inverse Fourier transformed to produce a sub-sampled, complex analytic time series restricted to that band~\cite{Pierini2025GFHv2}. This band spans from the initial frequency $f_0$ down to the value reached at the end of the observation window $T_{\rm obs}$ due to spin-down. This representation allows the pipeline to build the data segment needed for the analysis using the coherence time and observation time appropriate for each signal model.

The coherence time $T_{\rm FFT}$ is chosen so that the signal does not drift by more than one frequency bin during a single Fourier transform. Since the frequency evolution is fastest at the beginning of the signal, this condition is evaluated at the initial time. For the gravitational-wave-dominated spin-down model used in this work,
\begin{equation}
T_{\rm FFT} \simeq \frac{1}{\sqrt{|\dot{f}(t_0)|}} .
\label{eq:TFFT}
\end{equation}
This gives shorter FFTs for larger initial frequencies and larger ellipticities, where the spin-down is faster.

The observation time $T_{\rm obs}$ is chosen from the amplitude decay of the signal. For each pair $(f_0,\epsilon)$, we define $T_{\rm obs}$ as the time at which the instantaneous strain amplitude $h_0(t)$ has decreased to 
a fraction $1/\alpha_h$ of its initial value $h_0(t_0)$. The value of $\alpha_h$ is chosen for each $(f_0,\epsilon)$ pair by optimising the search sensitivity, following the procedure described in~\cite{2026PhRvD113l3042S}. The corresponding frequency band is then defined by the frequency evolution from the initial frequency $f_0$ down to the value reached at $T_{\rm obs}$.

After the data are prepared, the pipeline constructs a time-frequency peakmap by selecting local maxima in the equalized power spectrum~\cite{2005CQGra22S1255P}. The entire peakmap is then mapped into the transformed 
coordinates
\begin{equation}
x = \frac{1}{f^{n-1}},
\qquad
x_0 = \frac{1}{f_0^{n-1}},
\end{equation}
where $f$ is the gravitational-wave frequency at a given 
time, $f_0$ is the initial frequency and $n$ is the braking index. In this space, the power-law frequency evolution becomes linear,
\begin{equation}
x(t) = x_0 + (n-1)k(t-t_0).
\end{equation}
The Hough transform is then applied to transform the $(t,x)$ plane to the $(x_0,k)$ parameter space. A true signal is expected to produce an excess of counts in the Hough map at the parameters corresponding to its frequency evolution.

Candidates are selected from the Hough maps by dividing the parameter space into a grid of blocks in the $(x_0, k)$ plane. Within each block, the bin with the  highest Hough count is selected as the primary candidate, provided its count exceeds the locally estimated background median. A secondary candidate is additionally retained from the same block if it is sufficiently separated in $k$ from the primary one. Significance is measured by the critical ratio CR defined in Sec.~\ref{subsec:CR_estimation}. At this stage, candidates are selected according to their Hough number count relative to the local background; the CR is subsequently used as an additional selection criterion after the H1-L1 coincidence step. The block size and number of candidates per block were validated through injection studies on ER15 data, confirming that the procedure provides uniform coverage of the parameter space, significantly reducing the chances to select multiple candidates from the same noise fluctuation. Candidates from the two LIGO detectors are compared 
in the $(x_0, k)$ parameter space using a distance 
metric defined as
\begin{equation}
d_{\rm coin} = \sqrt{
\left(\frac{\Delta x_0}{\delta x_0}\right)^2 +
\left(\frac{\Delta k}{\delta k}\right)^2},
\label{eq:coinc_distance}
\end{equation}
where $\Delta x_0$ and $\Delta k$ are the differences 
in $x_0$ and $k$ between a candidate pair from H1 
and L1, and $\delta x_0$ and $\delta k$ are the grid spacings in the two parameters. A pair 
of candidates is considered coincident if 
$d_{\rm coin} \leq 2$. Only candidates that are coincident between H1 and L1 are kept for further analysis. 

\begin{figure}[ht]
\centering
\includegraphics[width=0.3\textwidth]{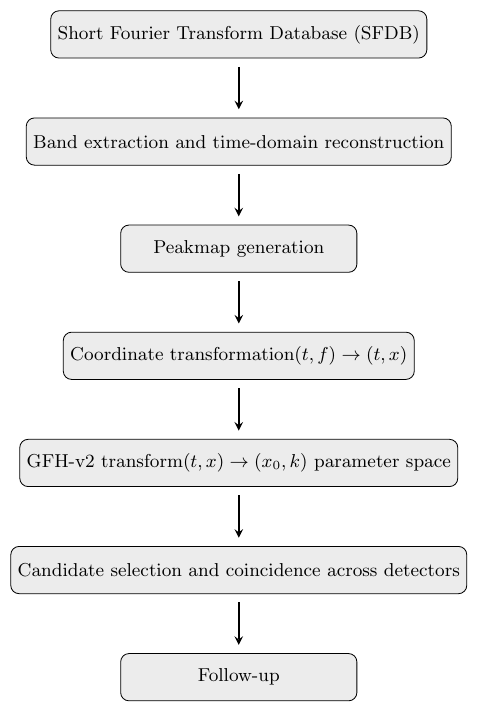}
\caption{Schematic overview of the main steps of the GFH-v2 search pipeline for a single search configuration $(f_0, \epsilon)$. The full analysis consists of an 
outer loop over all 700 configurations as described in Sec. \ref{sec:search_config}, each processed 
independently through these steps.}
\label{fig:workflow}
\end{figure}

The GFH-v2 workflow used in this search is shown in Fig.~\ref{fig:workflow}. The same pipeline is later used for the upper-limit injections, with the same search setup and candidate-selection procedure as in the real-data analysis.

\section{Search configuration and parameter space}
\label{sec:search_config}

We analyse the ER15 data described in Sec.~\ref{ER15data} over the frequency range $600$-$2000$~Hz. This range is motivated by rapidly rotating newborn neutron stars, for which the gravitational-wave emission is expected at high frequencies during the early spin-down phase~\cite{2020MNRAS4944838L,2007Ap&SS308119D}. The LIGO H1 and L1 datasets are analysed independently by the GFH-v2 pipeline, and their results are combined only at the coincidence stage, as described  in Sec.~\ref{subsec:gfhv2}.

The parameter space is described by the initial gravitational-wave frequency $f_0$, the ellipticity $\epsilon$, and the braking index $n$. The initial frequency is searched from $600$ to $2000$~Hz in steps of $10$~Hz. The ellipticity range is chosen as $
\epsilon \in [3\times10^{-4},3\times10^{-3}],$ and is divided into five sub-intervals. The braking index is fixed to $n=5$, corresponding to the gravitational-wave-dominated spin-down model used in this work. For each pair $(f_0,\epsilon)$, the coherence time $T_{\rm FFT}$, observation time $T_{\rm obs}$, and analysed frequency band are selected following the procedure described in~\cite{2026PhRvD113l3042S}. The coherence time is chosen so that the signal drift remains within one frequency bin during each Fourier transform (see Eq. \ref{eq:TFFT}). The observation time is set from the decay of the signal amplitude as described in Sec.~\ref{subsec:gfhv2}, keeping the part of the data where most of the signal contribution is expected.

The sky position is fixed to that of SN~2023ixf, using the electromagnetic localization of the source. The neutron star mass and radius are fixed to canonical values, $M=1.4\,M_\odot$ and $R=12$~km, which is used for the estimation of moment of inertia $I$, using the empirical relation~\cite{LATTIMER2016127}
\begin{equation}
I \approx MR^2\left(0.247 + 0.642\,\beta + 
0.466\,\beta^2\right),
\end{equation}
where $\beta = GM/c^2R$ is the compactness parameter. In total, the search consists of $140 \times 5 = 700$ independent configurations. Each configuration is processed separately by the GFH-v2 pipeline.

\begin{table}[t]
\centering
\caption{Search parameter space and fixed quantities used in the analysis.}
\label{tab:searchparams}
\begin{tabular}{lc}
\hline\hline
Parameter & Value / Range \\
\hline
Initial frequency, $f_0$ [Hz] & 600-2000, step 10 Hz \\
Ellipticity, $\epsilon$ & $3 \times 10^{-4}$-$3 \times 10^{-3}$, 5 grid steps \\
Braking index, $n$ & 5 \\
Sky position (RA, Dec) [deg] & (210.91, 54.31) \\
Neutron star mass, $M$ [$M_\odot$] & 1.4 \\
Neutron star radius, $R$ [km] & 12 \\
\hline\hline
\end{tabular}
\end{table}



\subsection{Critical ratio estimation}
\label{subsec:CR_estimation}

In previous work~\cite{2026PhRvD113l3042S}, the critical ratio (CR) was computed using the background statistics computed over the entire Hough map obtained from the peakmap, spanning the entire frequency range under consideration. While this approach provides a global characterization of the noise background, it assumes that the detector noise is statistically stationary across the analysed frequency range.

In the present analysis, the detector noise is treated as locally non-stationary. Variations in the detector noise spectral density together with narrow spectral disturbances produce changes in the background level across frequency. This effect becomes especially relevant in the context of the present search, where the analysis is performed over frequency bands that depend on the initial signal frequency $f_0$ and on the corresponding spin-down evolution. As a result, estimating the CR using statistics from the entire Hough map may not accurately represent the local noise background, which can affect the ranking of candidates.

To address this limitation, we adopt a local estimation of the noise statistics. For each 10 Hz search interval in $f_0$, the background statistics used for the CR are evaluated from the corresponding 10 Hz region of the Hough map in which candidates are selected. Thus, rather than estimating the background from the full Hough map, each candidate is assigned a CR using the noise statistics of its local 10 Hz search interval. The median and dispersion in this interval are evaluated using the robust estimators described in Appendix~\ref{app:robustCR}.

The critical ratio is thus defined as
\begin{equation}
    \mathrm{CR} = \frac{y - \mu_{\mathrm{loc}}}{\sigma_{\mathrm{loc}}},
    \label{eq:CR}
\end{equation}
where $y$ is the number count in a given Hough map bin, while $\mu_{\mathrm{loc}}$ and $\sigma_{\mathrm{loc}}$ denote the locally estimated background median and dispersion.

The validity of this approach has been assessed through injection studies performed on ER data. These tests demonstrate that the localized estimation improves the consistency of the CR across different frequency regions. We then determine a common threshold using the same injection-based false-alarm procedure described in~\cite{2026PhRvD113l3042S}, but using the local CR definition adopted here. Requiring a false-alarm probability below $0.1\%$ gives the threshold $\,CR_{\rm thr}=3.5\,$ which is used throughout the analysis. The CR threshold is applied after the H1-L1 coincidence step, only coincident candidates with $CR \geq CR_{\rm thr}$ are passed to the follow-up stage. Even though the search is performed over many independent search configurations, the CR threshold is primarily chosen to limit the number of candidates follow up stages, which provide the final distinction between noise fluctuations and astrophysical signals. 

\subsection{Computational implementation}

The analysis was carried out on the INFN CNAF computing cluster using the HTCondor batch system. The search was parallelized over the parameter space by assigning each frequency interval and ellipticity grid to an independent job. The 140 frequency intervals and five ellipticity sub-intervals give a total of 700 independent jobs.

The jobs were distributed across heterogeneous computing nodes of the CNAF cluster, with different processor architectures and computing performance. The performance of the allocated CPUs is characterized by the HEPSCORE benchmark; for the nodes used in this analysis, the HEPSCORE per core ranged from approximately 11 to 25. The execution time depends strongly on the search configuration, mainly because the coherence time, observation time, and analysed frequency band vary with $f_0$ and $\epsilon$.

\begin{figure*}[ht]
    \centering
    \includegraphics[scale=0.18]{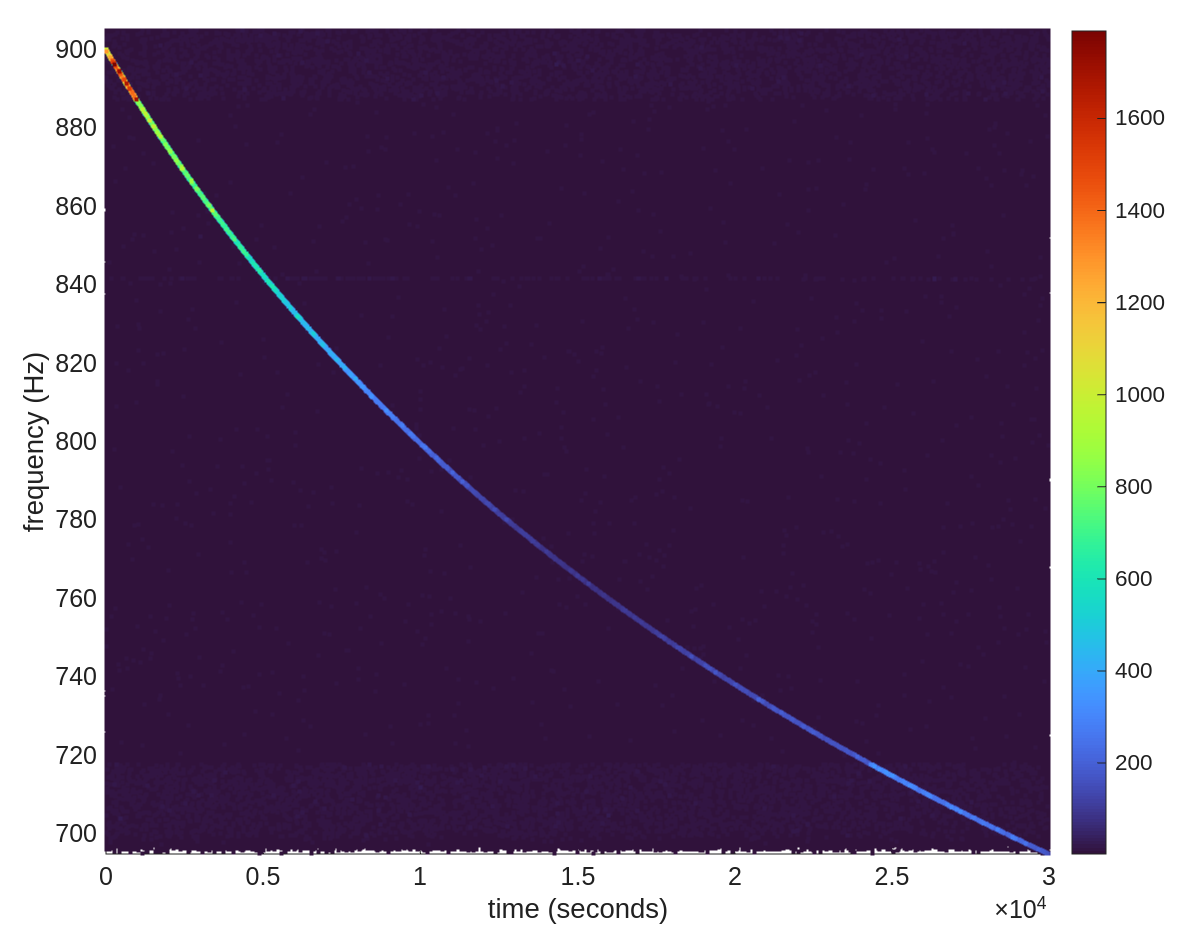}
    \begin{tikzpicture}
        \node[anchor=south west, inner sep=0] (main) at (0,0) {
            \includegraphics[scale=0.18]{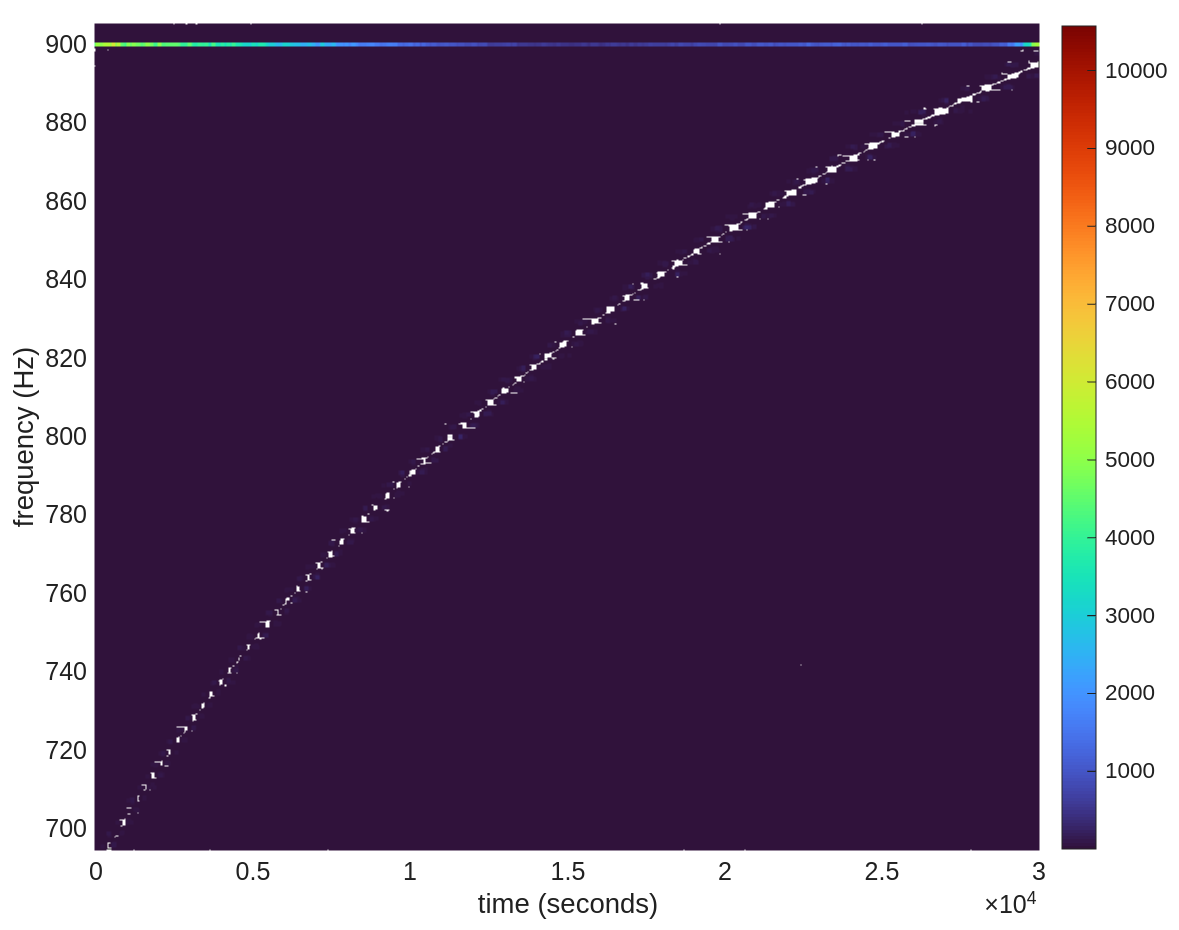}
        };

        \node at (4.2,1.85) {
            \includegraphics[scale=0.27]{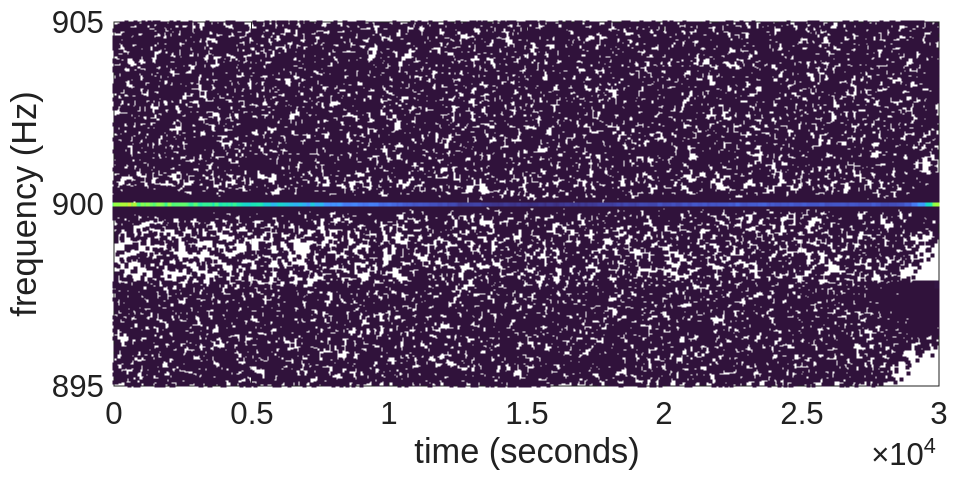}
        };
    \end{tikzpicture}
    \caption{Illustration of the heterodyne correction applied during the follow-up stage, shown for an injected signal in O4a data with $f_0 = 900$\,Hz, $\epsilon = 3\times10^{-3}$, and $d = 0.1$\,Mpc. Left: original time-frequency peakmap showing the power-law frequency evolution of the signal. Right: peakmap after applying the heterodyne correction using the recovered candidate parameters; the frequency evolution is largely removed, concentrating the signal into a nearly horizontal track near $f_0 = 900$\,Hz. The inset shows a zoomed view highlighting the concentration of signal power after the correction. The additional white curved feature in the corrected peakmap is produced by the heterodyne transformation of a nearly stationary spectral structure near the lower edge of the analyzed band and is unrelated to the injected signal.}
    \label{fig:heterodynedmaps}
\end{figure*}
The independent nature of the search configurations allows the jobs to be processed in parallel. Although the present analysis was performed using CPU resources, a GPU implementation of the GFH pipeline is also available and can be used to reduce the computational cost of future searches. The core GFH-v2 implementation is described in~\cite{Pierini2025GFHv2}.

\section{Search Results}
\label{sec:results}

The GFH-v2 search is run independently on the H1 and L1 data for each of the 700 configurations described in Sec.~\ref{sec:search_config}. Candidates are first selected from the Hough maps in each detector. A coincidence procedure is then applied between the two detectors in the $(x_0, k)$ parameter space, 
retaining only candidate pairs satisfying $d_{\rm coin} \leq 2$ (see Eq.~\eqref{eq:coinc_distance}). Only candidates that are consistent between H1 and L1 and have $CR \geq 3.5$ are kept.

After applying the coincidence requirement and the CR threshold, 131 coincident candidates remain. These candidates are the most significant outliers found by the search and are therefore passed to the follow-up stage. The coincident candidates are not distributed uniformly over the search configurations. This non-uniformity can arise from configuration-dependent differences in the analyzed bandwidth, observation time, coherence time, local spectral disturbances etc.

The 131 surviving candidates are therefore passed to the follow-up analysis described in the next section to determine whether they are consistent with astrophysical long-transient signals or with detector noise and instrumental artifacts.

\section{Candidate Follow-up}
\label{sec:followup}

The follow-up stage tests whether the coincident candidates are consistent with a real astrophysical long-transient signal. We use a procedure similar to that used in continuous-wave searches~\cite{2018PhRvD98j2004M} based on a preliminary heterodyne demodulation. For each candidate, the phase evolution expected from its recovered parameters $(f_0, k)$ is removed from the data through a heterodyne procedure~\cite{2019CQGra36a5008P}. The strain data $h(t)$ are multiplied by a complex exponential,
\begin{equation}
\tilde{h}(t) = h(t)\,e^{\,j\,\phi_{\rm sd}(t)},
\end{equation}
where the spin-down phase is obtained by integrating the frequency evolution of Eq.~\eqref{eq:frequency},
\begin{equation}
\phi_{\rm sd}(t) = 2\pi\int_{t_0}^{t}
\frac{f_0}{\left[1 + k(n-1)f_0^{n-1}
(t'-t_0)\right]^{\frac{1}{n-1}}}\,dt' + \phi_{t_0}.
\end{equation}
The expression above accounts for the intrinsic spin-down evolution. The Doppler modulation is corrected separately at the peakmap level by shifting the frequency bins according to the known sky position of SN~2023ixf. If the candidate is produced by a true signal, this correction should make the signal nearly monochromatic, allowing re-analysis with a longer coherence 
time and improved sensitivity.

To account for the finite resolution of the search grid, the heterodyne correction is not applied only at the exact recovered parameters but over a small grid of values centred on the candidate value. Specifically, for each candidate with recovered parameters $(f_{0,c},\, k_c)$, the correction is applied at each point of a $5\times5$ grid,
\begin{align}
f_0 &\in \{f_{0,c} + m\,\delta f_0 : m = -2,-1,0,1,2\}, \\
k   &\in \{k_c + l\,\delta k : l = -2,-1,0,1,2\},
\end{align}

where $\delta f_0$ and $\delta k$ are the bin widths of the Hough map in $f_0$ and $k$ respectively. This grid search ensures that any small offset between the true signal parameters and the recovered candidate parameters does not prevent the follow-up from detecting a real signal.

Figure~\ref{fig:heterodynedmaps} illustrates the effect of the heterodyne correction on an injected signal in O4a data. Before correction, the signal follows a power-law frequency evolution across the peakmap. After applying the heterodyne with the recovered candidate parameters, the frequency evolution is largely removed and the signal becomes nearly monochromatic confirming the expected behaviour.

After the phase correction, the data are analysed again with a longer coherence time. Since the phase evolution has been largely removed, the signal is expected to appear nearly monochromatic so that a standard Frequency Hough transform~\cite{2014PhRvD90d2002A} is applied at this stage. In the first follow-up stage we use
\begin{equation}
T_{\rm FFT,new}=3\,T_{\rm FFT,old}.
\end{equation}
A real signal would become more concentrated in frequency after the correction, leading to an increase in CR. This behaviour has been verified through injection studies performed on ER15 data, which confirm that real signals consistently show an increase in CR at this stage, while noise candidates do not. Candidates that do not show an increase in CR are thus rejected.

Candidates surviving the first stage would be followed by a second stage with
\begin{equation}
T_{\rm FFT,new}=6\,T_{\rm FFT,old}.
\end{equation}
This second step provides a stronger consistency check because a true signal should continue to gain significance when the coherence time is increased.

In this search, none of the 131 coincident candidates shows the expected increase in CR during the first follow-up stage. Therefore, no candidate is passed to the second stage. We conclude that all selected candidates are consistent with noise fluctuations or detector artifacts, and we find no evidence for tCW emission associated with SN~2023ixf.

\section{Upper Limit Estimation}
\label{sec:upperlimits}

Since no candidate survives the follow-up, we estimate upper limits on the maximum detectable distance. In this work, the upper limit is defined as the largest distance at which a signal with given source parameters would be recovered by the pipeline with at least 90\% efficiency.

For each frequency band and ellipticity interval, we first identify the maximum critical ratio, $CR_{\rm max}$, among the coincident candidates found in that region and excluded by the follow-up stage. Therefore, $CR_{\rm max}$ is determined separately for each search
configuration. Upper-limit injections are then performed separately for each of these configurations. Simulated signals with parameters belonging to the corresponding frequency band and ellipticity interval are added to the same ER15 data and analysed using the same configuration-specific search parameters as in the original search. The injections are repeated at increasing source distances. For each configuration, the upper limit is defined as the largest distance at which at least 90\% of the injected signals are recovered with $CR \geq CR_{\rm max}$ for that configuration and satisfy the H1-L1 coincidence criterion.

Upper limits are estimated only in regions of parameter space where coincident candidates are present in the original search. Configurations with no coincident candidates do not have a corresponding $CR_{\rm max}$ and no upper limit is assigned to them. This mainly affects the lower ellipticity intervals. In addition, for some low-ellipticity and low-frequency configurations where a coincident candidate is present, the injected signals do not reach 90\% recovery efficiency even at the minimum tested distance of $d_{\rm min}=0.1$\,Mpc. No upper limit is reported in either of these cases. The latter effect is consistent with the limited duration of the ER15 dataset: injection studies performed on longer O4a data stretches show that such slowly evolving signals become recoverable when more data are available.

\begin{figure}[ht]
\centering
\includegraphics[width=0.5\textwidth]{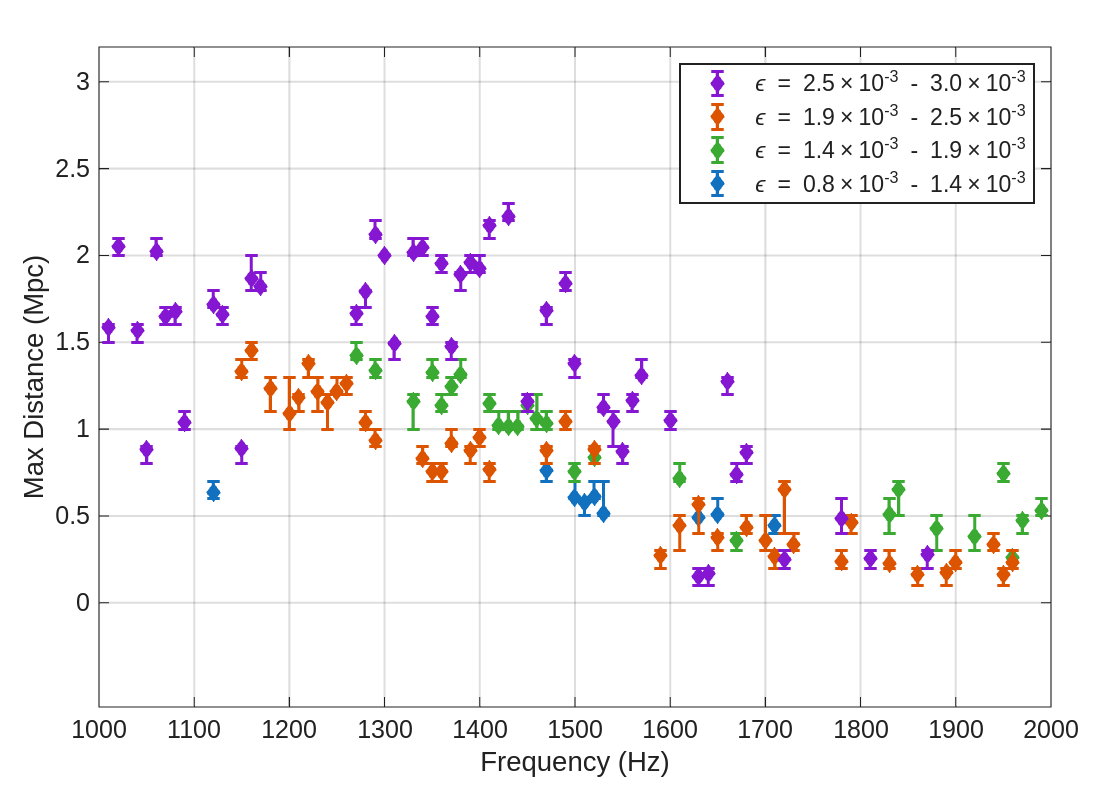}
\caption{Upper limits expressed as the maximum detectable distance (defined as in Ref.~\cite{2026PhRvD113l3042S}) as a function of the initial frequency $f_0$ for the different ellipticity intervals where an upper limit could be determined. Although five ellipticity intervals are included in the search, no upper limit could be determined for the lowest interval according to the procedure described in the text. Each color corresponds to a different ellipticity range.}
\label{fig:upperlimits}
\end{figure}
The resulting upper limits are shown in Fig.~\ref{fig:upperlimits}. Although five ellipticity intervals are included in the search, upper limits are reported only for four of them. For the lowest ellipticity interval, the conditions required to determine an upper limit are not satisfied over the analysed parameter space, either because no coincident candidate is available from the original search or because the 90\% recovery efficiency is not reached within the tested distance range. For higher ellipticities, upper limits are obtained across most of the analysed frequency range. For lower ellipticities, upper limits are available only in a restricted part of the parameter space because of the limited data duration and the smaller number of coincident candidates.
\begin{figure}[ht]
\centering
\includegraphics[width=0.5\textwidth]{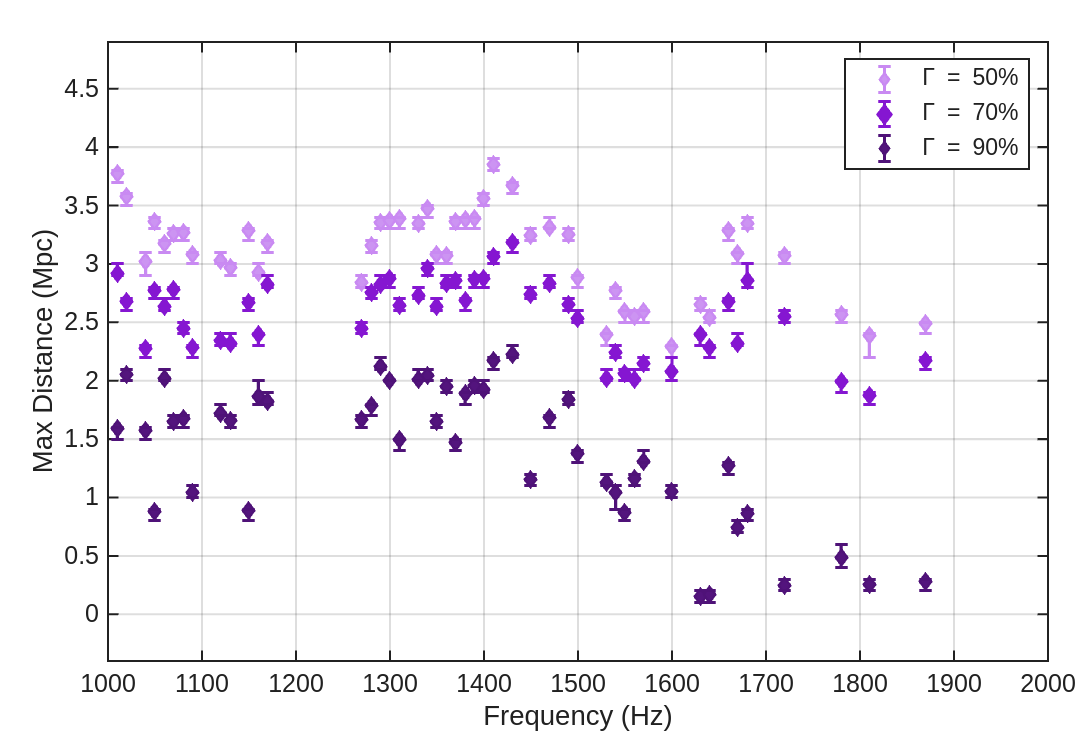}
\caption{Maximum detectable distance as a function of the initial frequency $f_0$ for the highest ellipticity interval $\epsilon \in [2.5\times10^{-3},\,3.0\times10^{-3}]$. Results are shown for recovery efficiencies of 50\%, 70\%, and 90\%.}
\label{fig:upperlimits_CL}
\end{figure}

Figure~\ref{fig:upperlimits_CL} shows the maximum detectable distance for the highest ellipticity interval, $\epsilon \in [2.5\times10^{-3},\,3.0\times10^{-3}]$ using three recovery efficiencies: 50\%, 70\%, and 90\%. As expected, the distance decreases when a higher recovery efficiency is required. We use the 90\% curve as the reference upper limit.

For the highest ellipticity interval, the 90\% upper limits are typically between about 1 and 2.5~Mpc over most of the analysed frequency range. The 50\% efficiency curve reaches distances up to about 4~Mpc. Although these distances are below the distance to M101 ($\sim6.8$~Mpc), the result provides a direct sensitivity estimate for long-transient searches in ER15 data and a useful benchmark for future searches with longer and more sensitive datasets.

\section{Conclusions}
\label{sec:conclusions}

We have performed a directed search for tCWs from the possible compact remnant of SN~2023ixf using LIGO ER15 data. The search was carried out with the GFH-v2 pipeline and covered initial gravitational-wave frequencies between 600 and 2000~Hz and ellipticities between $3\times10^{-4}$ and $3\times10^{-3}$.

After applying the coincidence requirement between the LIGO H1 and L1 detectors and a critical-ratio threshold of $CR_{\rm thr}=3.5$, 131 candidates were selected for follow-up. None of these candidates showed the expected increase in significance after the heterodyne follow-up. We therefore find no evidence for tCW emission associated with SN~2023ixf in the analysed ER15 data.

We placed upper limits on the maximum detectable distance as a function of initial frequency and ellipticity. For the highest ellipticity interval, the 90\% upper limits reach approximately 1-2.5~Mpc across most of the analysed frequency band. These limits are below the distance to M101, but they quantify the sensitivity achieved with ER15 data and show that GFH-v2 can be applied to real detector data in a directed search for a nearby core-collapse supernova.

Future searches using longer and more sensitive datasets from upcoming O5 and next-generation detectors will improve the detectable distance for these signals. The method used here can also be applied to future nearby magnetar-forming events identified by electromagnetic observations.

\begin{acknowledgments}
This work is partially supported by ICSC - Centro Nazionale di Ricerca in High Performance Computing, Big Data and Quantum Computing, funded by the European Union - NextGenerationEU.

This research has made use of data or software obtained from the Gravitational Wave Open Science Center (gw-openscience.org), a service of LIGO Laboratory, the LIGO Scientific Collaboration, the Virgo Collaboration, and KAGRA. LIGO Laboratory and Advanced LIGO are funded by the United States National Science Foundation (NSF) as well as the Science and Technology Facilities Council (STFC) of the United Kingdom, the Max-Planck-Society (MPS), and the State of Niedersachsen/Germany for support of the construction of Advanced LIGO and construction and operation of the GEO600 detector. Additional support for Advanced LIGO was provided by the Australian Research Council. 

Virgo is funded, through the European Gravitational Observatory (EGO), by the French Centre National de Recherche Scientifique (CNRS), the Italian Istituto Nazionale di Fisica Nucleare (INFN) and the Dutch Nikhef, with contributions by institutions from Belgium, Germany, Greece, Hungary, Ireland, Japan, Monaco, Poland, Portugal and Spain. 

The construction and operation of KAGRA are funded by Ministry of Education, Culture, Sports, Science and Technology (MEXT), and Japan Society for the Promotion of Science (JSPS), National Research Foundation (NRF) and Ministry of Science and ICT (MSIT) in Korea, Academia Sinica (AS) and the Ministry of Science and Technology (MoST) in Taiwan.

We also thank the Amaldi Research Center, for the clusters hosted in Rome INFN, where we have stored the LIGO/Virgo data used in this research and run part of the present analysis. We thank the INFN-CNAF computing staff for the resources we have used in this analysis and for their constant support. 

AC, DK, JRM, LM, OJP, RJ, and AS acknowledge support from the Universitat de les Illes Balears (UIB) through the Programa de Foment de la Recerca i la Innovació de la UIB 2024--2026, supported by the yearly plan of the Tourist Stay Tax (ITS2023-086); the Spanish Agencia Estatal de Investigación grants PID2022-138626NB-I00, PID2025-170644NA-I00, RED2024-153978-E and RED2024-153735-E, funded by MICIU/AEI/10.13039/501100011033 and the ERDF/EU; the Comunitat Autònoma de les Illes Balears through the Conselleria d'Educació i Universitats with funds from the ERDF (SINCO2022/18146, Plataforma HiTech-IAC3-BIO); and COST action SCALES CA24139. OJP additionally acknowledges support through the Spanish Ministerio de Ciencia, Innovación y Universidades Ramón y Cajal grant RYC2023-044489-I, funded by MCIN/AEI/10.13039/501100011033 and the FSE+ and cofinanced by UIB, and JRM is supported by the Spanish  Ministerio de Ciencia, Innovación y Universidades grant FPU22/01187.

AK and MB were partially supported by the Polish National Science Centre grant no. 2023/49/B/ST9/02777. 

KW is supported by the Australian Research Council Centre of Excellence for Gravitational Wave Discovery (OzGrav), project number CE230100016.

We further acknowledge Luca Rei for his constructive review and valuable feedback on the analysis presented in this work.

\end{acknowledgments}

\appendix

\section{Robust estimation of the local background}
\label{app:robustCR}

The local background statistics used to compute the critical ratio in Eq.~\eqref{eq:CR} are estimated using the robust statistical procedure described in Refs.~\cite{2009arXiv09052572F,2014PhRvD90d2002A}. Robust estimators are preferred over the conventional mean and standard deviation because they are much less affected by spectral disturbances and non-Gaussian tails, which are commonly present in GW detector data.

For each local Hough map, the background level is estimated by the median of the Hough counts,

\begin{equation}
\mu_{\rm loc}=\mathrm{median}(y),
\end{equation}

where $y$ denotes the Hough-map counts within the selected frequency interval.

The background dispersion is estimated using the robust estimator,

\begin{equation}
\sigma_{\rm loc}
=
\frac{\mathrm{median}\!\left(|y-\mu_{\rm loc}|\right)}
{0.6745},
\end{equation}

where the factor 0.6745 normalizes the estimator so that it is equal to the standard deviation for Gaussian-distributed data.

In the present analysis, these robust estimators are computed independently for each 10\,Hz search band. This local evaluation reduces the influence of narrow spectral artifacts and slow variations of the detector noise across frequency, providing a more reliable background estimate for candidate ranking. 

\bibliographystyle{apsrev4-2}
\bibliography{references}

\end{document}